\documentclass{article}

\usepackage{geometry}

\usepackage{epsfig}
\usepackage{amsfonts}

\usepackage{amssymb,amsmath, amsthm} 
\usepackage{dsfont}

\def\ve{\varepsilon}
\def\G{\Gamma}
\def\no{\nonumber}

\def\dis{\displaystyle}
\def\le{\left(}
\def\ri{\right)}

\begin{document}

\begin{titlepage}
\vskip 1cm
\begin{center}
{\Large \bf  Explicit calculation of multi-fold contour integrals \\
\vskip 2mm
of certain ratios of Euler gamma functions. Part 1.}\\
\vskip 5mm  
Ivan Gonzalez $^{(a)},$
Bernd A. Kniehl $^{(b)},$
Igor Kondrashuk $^{(c)},$ \\ 
\vskip 2mm
Eduardo A. Notte-Cuello $^{(d)},$  
Ivan Parra-Ferrada $^{(e)},$ 
Marko A. Rojas-Medar $^{(f)}$ \\
\vskip 1mm  
{\it  (a) Instituto de F\'isica y Astronom\'ia, Universidad de Valpara\'iso, \\ Av. Gran Breta\~na 1111,  Valpara\'iso, Chile}  \\
{\it  (b) II. Institut f\"ur Theoretische Physik, Universit\"at Hamburg,  \\ Luruper Chaussee 149,  22761 Hamburg, Germany} \\
{\it  (c) Grupo de Matem\'atica Aplicada {\rm \&} Grupo de F\'isica de Altas Energ\'ias, \\ 
          Departamento de Ciencias B\'asicas,  Universidad del B\'io-B\'io, Campus Fernando May, \\
          Av. Andres Bello 720, Casilla 447, Chill\'an, Chile} \\
{\it  (d) Departamento de Matem\'aticas, Facultad de Ciencias, Universidad de La Serena, \\
          Av. Cisternas 1200, La Serena, Chile}      \\
{\it  (e) Instituto de Matem\'atica y F\'isica, Universidad de Talca, \\ 2 Norte 685, Casilla 721, Talca, Chile } \\
{\it  (f) Instituto de Alta Investigaci\'on,  Universidad de Tarapac\'a,  \\ Casilla 7D, Arica, Chile}
\end{center}

\begin{abstract}
In this paper we proceed to study properties of Mellin-Barnes (MB) transforms of Usyukina-Davydychev (UD) functions.   
In our previous papers [Nuclear Physics B 870 (2013) 243], [Nuclear Physics B 876 (2013) 322] we showed that multi-fold Mellin-Barnes (MB) transforms of Usyukina-Davydychev (UD) 
functions may be reduced to two-fold MB transforms and that the   
higher-order UD functions were obtained in terms of a differential operator by applying  it to a slightly modified first UD
function.  The result is valid in $d=4$ dimensions and its analog 
in $d=4-2\ve$ dimensions exits  too [Theoretical and Mathematical Physics 177 (2013) 1515].   
In [Nuclear Physics B 870 (2013) 243] the chain of recurrent relations for analytically regularized UD functions 
was obtained implicitly by comparing  the left hand side and the right hand side of the diagrammatic relations between the diagrams with different loop orders. 
In turn, these  diagrammatic relations were obtained due to the method of loop reduction for the triangle ladder diagrams proposed in 1983 
by Belokurov and Usyukina. 
Here we reproduce these recurrent relations by calculating explicitly  via Barnes lemmas the contour integrals produced by the  left hand sides of the diagrammatic relations.  
In such a way we explicitly calculate a family of multi-fold contour integrals of certain ratios of Euler gamma functions.
We make a conjecture that similar results for the contour integrals are valid for a wider family of smooth functions which includes the MB transforms of UD functions.
\end{abstract}
\end{titlepage}

\section{Introduction}

Off-shell triangle-ladder and box-ladder diagrams are the only family of the Feynman diagrams which were calculated at any loop order, 
for example  in $d=4$ space-time dimensions \cite{Belokurov:1983km,Usyukina:1992jd,Usyukina:1993ch,Broadhurst:2010ds} with all indices equal to 1 in the momentum space representation (m.s.r.) and
in  $d=4-2\ve$ space-time dimensions with indices equal to $1-\ve$ on the rungs of ladders in the m.s.r. too \cite{Gonzalez:2012gu,Gonzalez:2012wk}.
For the important case of the ladder diagrams   with all indices equal to 1 in the m.s.r. in $d=4-2\ve$ space-time dimensions the on-shell result 
for this family of diagrams is known only at the first three loops in the form of expansion in terms of $\ve$ \cite{Smirnov,Bern:2005iz} up to a certain power of $\ve.$
The off-shell result for the whole family of the ladder diagrams  is unknown in $d=4-2\ve$ dimension.

The momentum integrals corresponding to the family of the ladder diagrams in $d=4$ space-time dimensions result in UD functions \cite{Usyukina:1992jd,Usyukina:1993ch}. The order of the UD function 
is the loop order in the ladder diagram \cite{Usyukina:1992jd,Usyukina:1993ch,Kondrashuk:2009us}. The ladder diagrams possess remarkable properties
at the diagrammatic level, for example, in Refs. \cite{Kondrashuk:2008ec,Kondrashuk:2008xq} it was shown that the UD functions are invariant with respect to Fourier 
transformations. In Ref. \cite{Allendes:2009bd,Kondrashuk:2009us} it has been shown that such a property of Fourier invariance may be generalized to any three-point 
Green function via Mellin-Barnes transformation.

MB transforms of the UD functions were investigated in Refs. \cite{Allendes:2012mr,Kniehl:2013dma}. It has been found under some analytical regularization of Ref. \cite{Belokurov:1983km} 
that MB transform of $n$-order UD function is a linear combination of MB transforms of three UD functions of $(n-1)$-order. This means any ladder diagram of this family may be reduced via a chain of recurrent relations 
to the one-loop scalar massless triangle diagram, which may be expressed for any indices and in any dimensions in terms of Appell function $F_4$ \cite{Boos:1990rg,Davydychev:1992xr}.  
This chain of the recurrent relations for the analytically regularized UD functions in the double-uniform limit when removing  this analytical regularization, 
is represented as a differential operator applied a to a slightly modified first UD function \cite{Kniehl:2013dma}. It has been shown there that if instead of MB transforms
of UD functions we write any smooth function of the same arguments the structure of this differential operator will be maintained the same in this double uniform limit. This operator will be 
applied to the function of the lowest order in this chain of recurrent relations.

However, in the present paper we show that in the particular case when in the integrand of the contour integrals
on the left hand sides of the diagrammatic relations the MB transforms of the UD functions stand, this chain of recurrent relations for the MB transforms of UD functions 
is produced by the contour integration. These contour integrals are calculated explicitly via the first and the second Barnes lemmas.  
Due to  observation done in the previous paragraph, we make a conjecture that similar results for the contour integrals are valid for a wider family of smooth functions written instead of 
MB transforms of UD functions. In the next papers we describe this family of functions and also describe what kind of changes should be made  for the contours of the integrals over complex variables 
for the case of other smooth functions different from certain ratios of Euler gamma functions. In this paper we focus on the contour integration via Barnes lemmas for the case 
when the integrand contains MB transforms of UD functions.

The Barnes lemmas were introduced in science about century ago. The first Barnes lemma has been proved in Ref. \cite{Barnes-1}, the second Barnes lemma has been proved in Ref. \cite{Barnes-2}.
They allow to integrate a product of several Euler gamma functions in a simple manner. The Barnes lemmas will help us to 
demonstrate the integral relations of Refs. \cite{Allendes:2012mr,Kniehl:2013dma} by doing complex integration along the contours typical for MB transformation. 
In Ref. \cite{Allendes:2012mr}  in order to obtain the results for the contour integrals  we simply  compared the left and the right parts 
of the diagrammatic relations.

\section{Proof}

The integral relation we need to prove via Barnes lemmas is Eq. (13) of Ref. \cite{Allendes:2012mr}, 
\begin{eqnarray} \label{R1}
\oint_C~dz_2dz_3~D^{(u,v)}[1+\ve_1-z_3,1+\ve_2-z_2,1+\ve_3] 
D^{(z_2,z_3)}[1+\ve_2,1+\ve_1,1+\ve_3] =  \no\\
J\left[ \frac{D^{(u,v-\ve_2)}[1-\ve_1]}{\ve_2\ve_3}
+  \frac{D^{(u,v)}[1+\ve_3]}{\ve_1\ve_2}  + \frac{ D^{(u-\ve_1,v)}[1-\ve_2]}{\ve_1\ve_3}  \right]  
\end{eqnarray}
in which the parameters $\ve_1,\ve_2$ and $\ve_3$ are three complex variables of analytical regularization used in Ref. \cite{Belokurov:1983km}, subject to condition 
\begin{eqnarray*}
\ve_1 + \ve_2 + \ve_3 = 0, 
\end{eqnarray*}
\begin{figure}[hb!] 
\centering\includegraphics[scale=0.83]{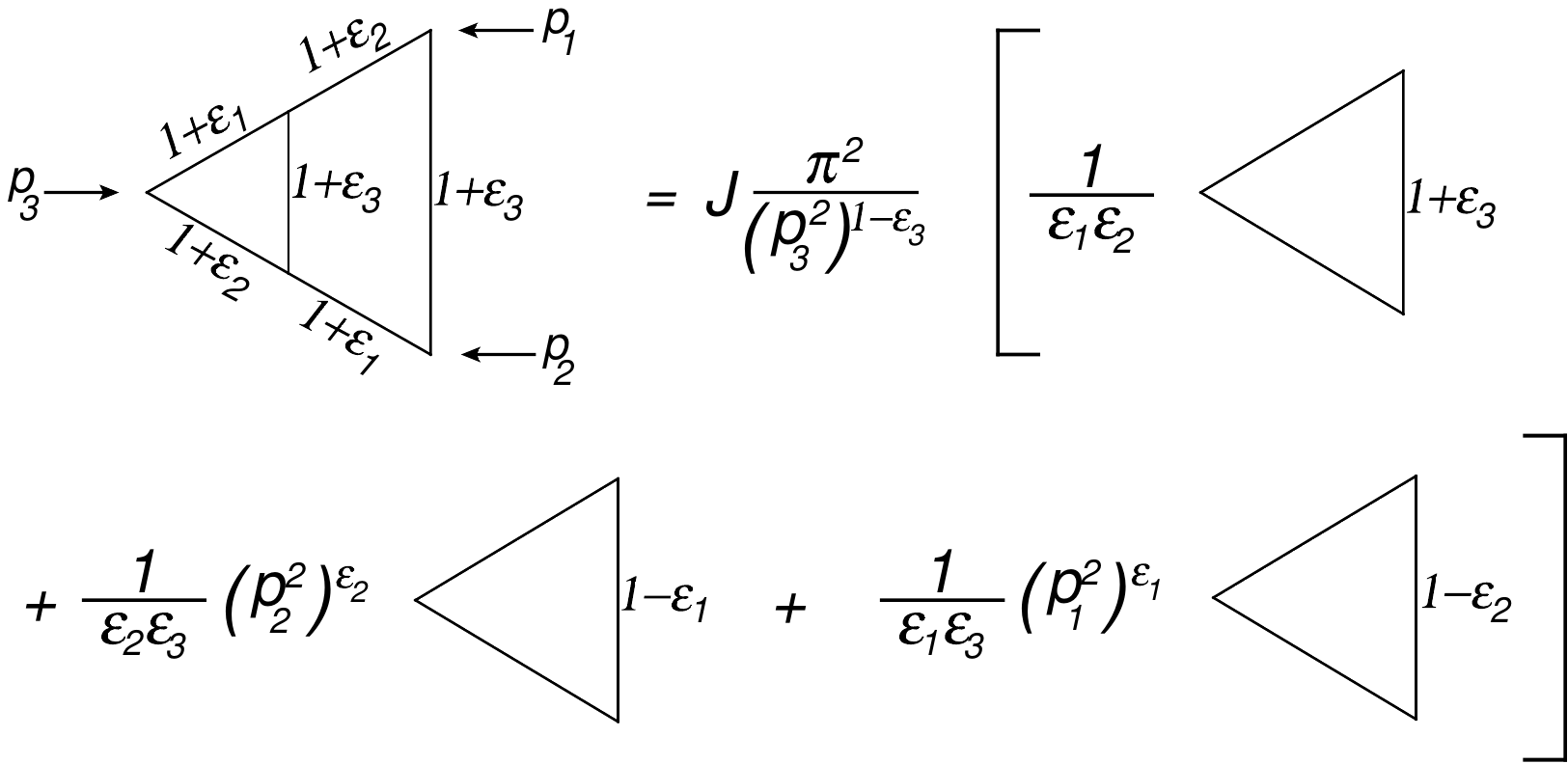}
\caption{\footnotesize  Equation (25) of Ref.\ \cite{Usyukina:1992jd} is the origin of integral relation Eq.(\ref{R1}).}
\label{figure-8}
\end{figure}
the factor $J$ is a ratio of Euler gamma functions  
\begin{eqnarray*}
J =   \frac{\G(1-\ve_1)\G(1-\ve_2) \G(1-\ve_3) }{\G(1+\ve_1)\G(1+\ve_2) \G(1+\ve_3) }, 
\end{eqnarray*}
the function $D^{(z_2,z_3)}[\nu_1,\nu_2,\nu_3]$ is the MB transform of one-loop triangle integral in the momentum space $J(\nu_1,\nu_2,\nu_3)$ taken in Ref.\cite{Allendes:2012mr} from
Refs. \cite{Davydychev:1992xr,Usyukina:1992jd,Usyukina:1993ch}, 
\begin{eqnarray} \label{R2}
D^{(z_2,z_3)}[\nu_1,\nu_2,\nu_3] = \frac{ \G \le -z_2 \ri \G \le -z_3 \ri \G \le -z_2 -\nu_2-\nu_3 + d/2 \ri 
\G \le -z_3-\nu_1-\nu_3 + d/2 \ri }
{\Pi_{i} \G(\nu_i) } \no\\
\times 
\frac{ \G \le z_2 + z_3  + \nu_3 \ri  \G \le  \Sigma \nu_i - d/2 + z_3 + z_2 \ri }
{\G(d-\Sigma_i \nu_i)},
\end{eqnarray} 
and for the brevity the notation 
\begin{eqnarray} \label{R3}
 D^{(u,v)}[1+\nu] \equiv  D^{(u,v)}[1,1,1+\nu]
\end{eqnarray}
is used.  The integral relations in Eq. (\ref{R1}) is produced by the diagrammatic relation between scalar Feynman diagrams in the momentum space  given in Fig. \ref{figure-8} which is 
Eq. (25) of Ref. \cite{Usyukina:1992jd}.  This graphical equation was originally obtained by using the uniqueness relations from Ref. \cite{Usyukina:1992jd},
and, in particular, this is why the analytic regularization indices were chosen in such a way that $\ve_1 + \ve_2 + \ve_3=0.$  
The derivation of this diagrammatic relation is reviewed in details in Ref. \cite{Allendes:2012mr}. Also, in  Ref. \cite{Allendes:2012mr} 
the derivation of Eq. (\ref{R1}) from the diagrammatic relation of Fig. \ref{figure-8} may be found.

According to Eqs. (\ref{R2}) and  (\ref{R3}), we write 
\begin{eqnarray*}
D^{(z_2,z_3)}[1+\ve_2,1+\ve_1,1+\ve_3] = \\
\frac{ \G \le -z_2 \ri \G \le -z_3 \ri \G \le -z_2 +\ve_2\ri \G \le -z_3 +\ve_1\ri   \G \le 1+z_2 +z_3\ri  \G \le 1+z_2 +z_3 +\ve_3\ri}
{\G \le 1+\ve_1\ri \G \le 1+\ve_2 \ri \G \le 1+\ve_3\ri  },  \no\\
D^{(u,v)}[1+\ve_1 -z_3,1+\ve_2-z_2,1+\ve_3] =  \no \\
\frac{ \G \le -u \ri \G \le -v \ri \G \le -u +\ve_1 +z_2\ri \G \le -v +\ve_2 +z_3\ri  \G\le 1+u+v+\ve_3\ri \G \le 1-z_2 -z_3 + u + v\ri} 
{\G \le 1+\ve_1 -z_3\ri \G \le 1+\ve_2 -z_2\ri \G \le 1+\ve_3\ri\G \le 1+z_2+z_3\ri  }, 
\end{eqnarray*}
and the integrand on the left hand side of Eq.(\ref{R1}) is 
\begin{eqnarray} \label{IM}
D^{(u,v)}[1+\ve_1 -z_3,1+\ve_2-z_2,1+\ve_3]D^{(z_2,z_3)}[1+\ve_2,1+\ve_1,1+\ve_3]  = \no\\
\frac{ \G \le -u \ri \G \le -v \ri \G \le 1+u +v+\ve_3\ri }{\G \le 1+\ve_1\ri \G \le 1+\ve_2\ri 
\G^2 \le 1+\ve_3\ri}  \times \no\\
\times \frac{\G \le -z_2 \ri \G \le -z_3 \ri\G \le 1+z_2 + z_3 + \ve_3 \ri \G \le -u + \ve_1 +z_2\ri \G \le -v + \ve_2 +z_3\ri  \G \le 1-z_2 -z_3 +u+v\ri}
{(\ve_1-z_3)(\ve_2-z_2)}.
\end{eqnarray}
On the right hand side of Eq.(\ref{R1}) we should obtain  
\begin{eqnarray}
\frac{D^{(u,v)}[1+\ve_3]}{\ve_1\ve_2} &=& 
\frac{1}{\ve_1\ve_2}\frac{\G(-u)\G(-v)\G(-\ve_3 -u)\G(-\ve_3 -v)\G^2(1+\ve_3 + u + v)}{\G(1-\ve_3)\G(1+\ve_3)},  \label{T1}\\
\frac{D^{(u,v-\ve_2)}[1-\ve_1]}{\ve_2\ve_3} &=& 
\frac{1}{\ve_2\ve_3}\frac{\G(-u)\G(\ve_2-v)\G(\ve_1-u)\G(-\ve_3  -v)\G^2(1+\ve_3 + u + v)}{\G(1-\ve_1)\G(1+\ve_1)}, \label{T2}\\
\frac{ D^{(u-\ve_1,v)}[1-\ve_2]}{\ve_1\ve_3} &=& 
\frac{1}{\ve_1\ve_3}\frac{\G(\ve_1-u)\G(-v)\G(-\ve_3-u)\G(\ve_2 -v)\G^2(1+\ve_3 + u + v)}{\G(1-\ve_2)\G(1+\ve_2)}.  \label{T3}
\end{eqnarray}

The poles at the points $z_2 =\ve_2$ and $z_3 =\ve_1$ were originally ``right'' since they come from the Euler gamma functions 
with negative signs of the integration variables of their arguments. The contribution of the corresponding residues at the points 
$z_2 =\ve_2$ and $z_3 =\ve_1$ in the integrand of  Eq.(\ref{R1}) which is Eq.(\ref{IM}) reproduces term (\ref{T1}) on the right hand side 
of Eq.(\ref{R1}). 

To obtain terms (\ref{T2}) and (\ref{T3}) on the right hand side of Eq.(\ref{R1}) we need to use the Barnes lemmas. The first lemma has been published in 1908 in Ref. \cite{Barnes-1}
\begin{eqnarray} \label{Barnes-1}
\oint_C~dz ~\G \le \lambda_1 + z \ri  \G \le \lambda_2 + z \ri \G \le \lambda_3 - z \ri \G \le \lambda_4 - z \ri   =  \no\\
\frac{\G \le \lambda_1 + \lambda_3 \ri \G \le \lambda_1 + \lambda_4 \ri \G \le \lambda_2 + \lambda_3 \ri \G \le \lambda_2 + \lambda_4 \ri}
{\G \le \lambda_1 + \lambda_2 + \lambda_3 + \lambda_4  \ri },  
\end{eqnarray}
in which  $\lambda_1, \lambda_2, \lambda_3, \lambda_4$ are complex numbers, chosen in a such a way  that on the right hand side of Eq. (\ref{Barnes-1}) 
there are no singularities, while the second Barnes lemma has been published in 1910 in  Ref. \cite{Barnes-2}, 
\begin{eqnarray} \label{Barnes-2}
\oint_C~dz ~\frac{\G \le \lambda_1 + z \ri  \G \le \lambda_2 + z \ri \G \le \lambda_3 + z \ri \G \le \lambda_4 - z \ri \G \le \lambda_5 - z \ri}
{\G \le \lambda_1 + \lambda_2 + \lambda_3 + \lambda_4 + \lambda_5 + z \ri }  =  \no\\
 \frac{\G \le \lambda_1 + \lambda_4 \ri \G \le \lambda_2 + \lambda_4 \ri \G \le \lambda_3 + \lambda_4 \ri \G \le \lambda_1 + \lambda_5 \ri 
\G \le \lambda_2 + \lambda_5 \ri  \G \le \lambda_3 + \lambda_5 \ri   }
{\G \le \lambda_1 + \lambda_2 + \lambda_4 + \lambda_5  \ri \G \le \lambda_1 + \lambda_3 + \lambda_4 + \lambda_5  \ri 
\G \le \lambda_2 + \lambda_3 + \lambda_4 + \lambda_5  \ri} 
\end{eqnarray}
in which  $\lambda_1, \lambda_2, \lambda_3, \lambda_4,  \lambda_5$ are complex numbers, chosen in a such a way  that on the right hand side of Eq. (\ref{Barnes-2}) 
there are no singularities.

The integrand of Eq. (\ref{IM}) may be represented as 
\begin{eqnarray} \label{imm}
\frac{1}{z_3-\ve_1}\frac{1}{z_2-\ve_2} \G \le -z_2 \ri \G \le -z_3 \ri \G \le 1+z_2 + z_3 + \ve_3 \ri \G \le -u + \ve_1 +z_2\ri \times \no\\
\times \G \le -v + \ve_2 +z_3\ri \G \le 1-z_2 -z_3 +u+v\ri = \no\\
= \frac{z_2 + z_3 + \ve_3}{(z_3-\ve_1)(z_2-\ve_2)} \G \le -z_2 \ri \G \le -z_3 \ri \G \le z_2 + z_3 + \ve_3 \ri \G \le -u + \ve_1 +z_2\ri \times \no\\
\times \G \le -v + \ve_2 +z_3\ri \G \le 1-z_2 -z_3 +u+v\ri = \no\\
= \le \frac{1}{z_3-\ve_1} +  \frac{1}{z_2-\ve_2} \ri \G \le -z_2 \ri \G \le -z_3 \ri \G \le z_2 + z_3 + \ve_3 \ri 
\G \le -u + \ve_1 +z_2\ri \times \no\\
\times \G \le -v + \ve_2 +z_3\ri \G \le 1-z_2 -z_3 +u+v\ri,
\end{eqnarray}
and this is a sum of two terms. We consider the second term, 
\begin{eqnarray*}
\oint_C~dz_2\frac{1}{z_2-\ve_2}  \G \le -z_2 \ri \G \le -u + \ve_1 +z_2\ri    \oint_C~dz_3 \G \le -z_3 \ri \G \le z_2 + z_3 + \ve_3 \ri \times \\
\G \le -v + \ve_2 +z_3\ri \G \le 1-z_2 -z_3 +u+v\ri,
\end{eqnarray*}
in which the integral over $z_3$  may be calculated via the first Barnes lemma,   
\begin{eqnarray*}
\oint_C~dz_2\frac{1}{z_2-\ve_2}  \G \le -z_2 \ri \G \le -u + \ve_1 +z_2\ri    \oint_C~dz_3 \G \le -z_3 \ri \G \le z_2 + z_3 + \ve_3 \ri \times \\
\G \le -v + \ve_2 +z_3\ri \G \le 1-z_2 -z_3 +u+v\ri = \\
\oint_C~dz_2\frac{1}{z_2-\ve_2}\G \le -z_2 \ri \G \le -u + \ve_1 +z_2\ri \times \\
\times \frac{\G \le z_2 +\ve_3 \ri  \G \le -v +\ve_2 \ri \G\le 1 + \ve_3 + u +v \ri \G\le 1 + \ve_2 + u -z_2\ri } 
{\G \le 1+u -\ve_1\ri} =  \\
= \frac{\G \le -v +\ve_2 \ri \G\le 1 + \ve_3 + u +v \ri}{\G \le 1+u -\ve_1\ri}  \oint_C~dz_2\frac{1}{z_2-\ve_2}
\G \le z_2 +\ve_3 \ri  \G \le -z_2 \ri \times \\
\times \G\le -u + \ve_1  + z_2 \ri \G\le 1 + \ve_2 + u -z_2\ri.
\end{eqnarray*}

Now we do reflection of the complex variable $z_2$ of contour integration,  $z_2 \longrightarrow -z_2,$ and apply the second Barnes lemma,  
\begin{eqnarray*}
\frac{\G \le -v +\ve_2 \ri \G\le 1 + \ve_3 + u +v \ri}{\G \le 1+u -\ve_1\ri}  \oint_C~dz_2\frac{1}{z_2-\ve_2}
\G \le z_2 +\ve_3 \ri  \G \le -z_2 \ri \times \\
\times \G\le -u + \ve_1  + z_2 \ri \G\le 1 + \ve_2 + u -z_2\ri = \\
- \frac{\G \le -v +\ve_2 \ri \G\le 1 + \ve_3 + u +v \ri}{\G \le 1+u -\ve_1\ri}  \oint_C~dz_2\frac{1}{z_2+\ve_2}
\G \le -z_2 +\ve_3 \ri  \G \le z_2 \ri \times \\
\times \G\le -u + \ve_1  - z_2 \ri \G\le 1 + \ve_2 + u + z_2\ri = \\
- \frac{\G \le -v +\ve_2 \ri \G\le 1 + \ve_3 + u +v \ri}{\G \le 1+u -\ve_1\ri}  \oint_C~dz_2\frac{\G(z_2+\ve_2)}{\G(1+z_2+\ve_2)}
\G \le -z_2 +\ve_3 \ri  \G \le z_2 \ri \times \\
\times \G\le -u + \ve_1  - z_2 \ri \G\le 1 + \ve_2 + u + z_2\ri = \\
- \frac{\G \le -v +\ve_2 \ri \G\le 1 + \ve_3 + u +v \ri}{\G \le 1+u -\ve_1\ri}  \G(\ve_3)\G(-\ve_1)\G \le 1 + u - \ve_1 \ri \times \\
\frac{\G \le -u - \ve_3 \ri  \G \le -u + \ve_1 \ri \G\le 1 - \ve_3 \ri}{ \G\le 1 + \ve_2 \ri\G\le -u \ri  } = \\
\frac{1}{\ve_1\ve_3} \frac{\G(1-\ve_1)\G(1+\ve_3)\G(1-\ve_3) }{\G\le 1 + \ve_2 \ri} \times\\
\frac{\G \le -v +\ve_2 \ri \G\le 1 + \ve_3 + u +v \ri\G \le -u - \ve_3 \ri  \G \le -u + \ve_1 \ri }{ \G\le -u \ri  }.
\end{eqnarray*}

Taking into account the factor from Eq. (\ref{IM}), we obtain 
\begin{eqnarray*}
\frac{ \G \le -u \ri \G \le -v \ri \G \le 1+u +v+\ve_3\ri }{\G \le 1+\ve_1\ri \G \le 1+\ve_2\ri \G^2 \le 1+\ve_3\ri} \times 
\frac{1}{\ve_1\ve_3} \frac{\G(1-\ve_1)\G(1+\ve_3)\G(1-\ve_3) }{\G\le 1 + \ve_2 \ri} \times\\
\frac{\G \le -v +\ve_2 \ri \G\le 1 + \ve_3 + u +v \ri\G \le -u - \ve_3 \ri  \G \le -u + \ve_1 \ri }{ \G\le -u \ri  } = \frac{J}{\ve_1\ve_3} D^{(u-\ve_1,v)}[1-\ve_2].
\end{eqnarray*}
The first term in Eq. (\ref{imm}) analogously reproduces term $\dis{\frac{J}{\ve_2\ve_3} D^{(u,v-\ve_2)}[1-\ve_1]}$ on the right hand side of Eq. (\ref{R1}). 
We need to comment that there is no double counting residues at the points $z_2 =\ve_2$ and $z_3 =\ve_1$ because after the reflection these points become ``left'' 
poles, that is, they come from Euler gamma functions with positive signs of the integration variable in the arguments of gamma functions, 
while we calculate the ``right'' residues only, that is, the residues which come from   Euler gamma functions with negative signs of the integration variable 
in the arguments of gamma functions.

\section{Conclusion}

We showed in Ref.\cite{Kniehl:2013dma} [Nuclear Physics B 876 (2013) 322]  that structure of the chain of  recurrent relations for the Mellin-Barnes transforms  of the analytically regularized UD functions guarantees 
the finiteness of the double-uniform limit when removing the analytical regularization. 
The limit was expressed in terms of a differential operator. This operator is the same for any smooth function written instead of the MB transforms of the UD functions 
and has nothing to do with explicit form of these MB transforms.  The present paper shows that the first and the second 
Barnes lemmas permit to work out the contour integration only in a particular case of MB transforms of UD functions to produce this chain of the recurrent relations. 
For a wider family of smooth functions the Barnes lemmas should be replaced with another integration trick by using  more complicate contour of integration.

\subsection*{Acknowledgments}

The work of B.A.K. was supported in part by the German Science Foundation (DFG)
within the Collaborative Research Center SFB 676 ``Particles, Strings and the
Early Universe'' and by the German Federal Ministry for Education and Research
(BMBF) through Grant No.\ 05H12GUE.
The work of I.K. was supported in part 
by Fondecyt (Chile) Grants Nos. 1040368, 1050512 and 1121030, by DIUBB (Chile) Grant Nos.  125009,  GI 153209/C  and GI 152606/VC.  
Also, the work of I.K. is supported by Universidad del B\'\i o-B\'\i o and Ministerio de Educacion (Chile)
within Project No.\ MECESUP UBB0704-PD018.
He is grateful to the Physics Faculty of Bielefeld University for accepting
him as a visiting scientist and for the kind hospitality and the excellent
working conditions during his stay in Bielefeld. 
E. Notte-Cuello's  work was partially supported by project DIULS PR15151, Universidad de La Serena.
The work of I.P.F. was supported in part by Fondecyt (Chile) Grant No.\
1121030 and by Beca Conicyt (Chile) via Master fellowship CONICYT-PCHA/Magister Nacional/2013-22131319. 
The work of M.R.M. was supported in part by Project No.\ MTM2012-32325,
by Ministerio de Ciencia e Innovaci\'on, Espa\~na,
by Fondecyt (Chile) Grant Nos.\ 1080628 and 1120260,
and by DIUBB (Chile) Grant No.\ 121909 GI/C-UBB. 
This paper is based on the talk of I. K. at XVIII International Congress on Mathematical Physics, Santiago de Chile, July 27 - August 1, 2015, 
and he is grateful to Rafael Benguria for inviting him.


\begin{thebibliography}{99}


\bibitem{Belokurov:1983km}
V.~V.~Belokurov and N.~I.~Ussyukina,
``Calculation of ladder diagrams in arbitrary order,''
J.\ Phys.\ A: Math.\ Gen.\ {\bf 16} (1983) 2811.

\bibitem{Usyukina:1992jd}
 N.~I.~Ussyukina and A.~I.~Davydychev,
``An approach to the evaluation of three- and four-point ladder diagrams,''
 Phys.\ Lett.\  B {\bf 298} (1993) 363.

 
\bibitem{Usyukina:1993ch}
N.~I.~Ussyukina and A.~I.~Davydychev,
``Exact results for three- and four-point ladder diagrams with an arbitrary
number of rungs,''
Phys.\ Lett.\  B {\bf 305} (1993) 136.
  
  

\bibitem{Broadhurst:2010ds}
D.~J.~Broadhurst and A.~I.~Davydychev,
``Exponential suppression with four legs and an infinity of loops,''
Nucl.\ Phys.\ B (Proc.\ Suppl.) {\bf 205--206} (2010) 326
[arXiv:1007.0237 [hep-th]].

  
  
\bibitem{Gonzalez:2012gu}
  I.~Gonzalez and I.~Kondrashuk,
  ``Belokurov-Usyukina loop reduction in non-integer dimension,''
  Phys.\ Part.\ Nucl.\  {\bf 44} (2013) 268
  [arXiv:1206.4763 [hep-th]].
  
  
  
\bibitem{Gonzalez:2012wk}
  I.~Gonzalez and I.~Kondrashuk,
  ``Box ladders in a noninteger dimension,''
  Theor.\ Math.\ Phys.\  {\bf 177} (2013) 1515
   [Teor.\ Mat.\ Fiz.\  {\bf 177} (2013) no.1,  276]
  [arXiv:1210.2243 [hep-th]].
  
  
 

\bibitem{Smirnov}
V.~A.~Smirnov,
``Evaluating Feynman integrals,''
Springer Tracts Mod.\ Phys.\ {\bf 211} (2004) 1. 


\bibitem{Bern:2005iz}
Z.~Bern, L.~J.~Dixon and V.~A.~Smirnov, 
``Iteration of planar amplitudes in maximally supersymmetric Yang-Mills theory at three loops and beyond,''
  Phys.\ Rev.\  D {\bf 72} (2005) 085001
  [hep-th/0505205].

  
  
\bibitem{Kondrashuk:2009us}
I.~Kondrashuk and A.~Vergara, 
``Transformations of triangle ladder diagrams,'' 
JHEP {\bf 1003} (2010) 051
[arXiv:0911.1979 [hep-th]].
  
  

\bibitem{Kondrashuk:2008ec} 
I.~Kondrashuk and A.~Kotikov,
``Fourier transforms of UD integrals,'' 
in {\it Analysis and Mathematical Physics}, 
Birkh\"auser Book Series Trends in Mathematics,
edited by B.~Gustafsson and A.~Vasil'ev,
(Birkh\"auser, Basel, Switzerland, 2009), pp.~337
[arXiv:0802.3468 [hep-th]].

\bibitem{Kondrashuk:2008xq}
I.~Kondrashuk and A.~Kotikov,
``Triangle UD integrals in the position space,''  
JHEP {\bf 0808} (2008) 106
[arXiv:0803.3420 [hep-th]].



\bibitem{Allendes:2009bd}
P.~Allendes, N.~Guerrero, I.~Kondrashuk and E.~A.~Notte Cuello, 
``New four-dimensional integrals by Mellin-Barnes transform,'' 
J.\ Math.\ Phys.\ {\bf 51} (2010) 052304
[arXiv:0910.4805 [hep-th]]. 

\bibitem{Allendes:2012mr}
P.~Allendes, B.~A.~Kniehl, I.~Kondrashuk, E.~A.~Notte-Cuello and M.~Rojas-Medar,
``Solution to Bethe-Salpeter equation via Mellin-Barnes transform,''
  Nucl.\ Phys.\ B {\bf 870} (2013) 243
  [arXiv:1205.6257 [hep-th]].


\bibitem{Kniehl:2013dma}
  B.~A.~Kniehl, I.~Kondrashuk, E.~A.~Notte-Cuello, I.~Parra-Ferrada and M.~Rojas-Medar,
  ``Two-fold Mellin-Barnes transforms of Usyukina-Davydychev functions,''
  Nucl.\ Phys.\ B {\bf 876} (2013) 322
  [arXiv:1304.3004 [hep-th]].
  
  
\bibitem{Boos:1990rg}
E.~E.~Boos and A.~I.~Davydychev,
``A Method of evaluating massive Feynman integrals,''
Teor.\ Mat.\ Fiz.\  {\bf 89} (1991) 56
[Theor.\ Math.\ Phys.\  {\bf 89} (1991) 1052].

\bibitem{Davydychev:1992xr}
A.~I.~Davydychev,
``Recursive algorithm of evaluating vertex-type Feynman integrals,''
J.\ Phys.\ A: Math.\ Gen.\ {\bf 25} (1992) 5587.
  
  
\bibitem{Barnes-1} 
E. W. Barnes, 
``A new development of the theory of the hypergeometric functions,'' 
Proceedings of the London Mathematical Society. Second Series, 6 (1908) 141 


\bibitem{Barnes-2} 
E. W. Barnes, 
``A transformation of generalized hypergeometric series,'' 
The Quarterly Journal of Pure and Applied Mathematics, 41 (1910) 136



\end{thebibliography}
\end{document}